\documentstyle[aps,prl,multicol,epsf]{revtex}
\begin{document}
\author{W. Wu $^1$, W. Wang$^2$ ,X. X. Yi$^{1}$}
\address{ $^1$Institute of Theoretical Physics, Northeast Normal University, Changchun 130024, China\\
 $^2$ Department of Telecom Engineering, Changchun Institute of
Posts and Telecommunications at Jilin University, Changchun
130012, China}
\title{Dynamics of Distillability }
\maketitle
\begin{abstract}
The time evolution of a maximally entangled bipartite systems is
presented in this paper. The distillability criterion is given in
terms of Kraus operators. Using the criterion, we discuss the
distillability of $2\times 2$ and $n\times n (n>2)$  systems in
their evolution process. There are two distinguished processes,
dissipation and decoherence, which may destroy the distillability.
We discuss the effects of those processes on distillability in
details.

{\bf PACS number(s): 03.67.-a, 03.65.Bz}
\end{abstract}
\vspace{4mm}
\begin{multicols}{2}[]
 Maximally entangled states are an essential ingredient in most
applications of quantum information[1,2]. In particular, in
quantum communications, by using entangled states several
proposals were devised to transmit secret messages between two
locally separated parties[3-5]. These proposals have been
successfully implemented experimentally by using pairs of photons
generated via parametric down conversion[6-9]. On the other hand,
quantum entanglement is a fragile feature, which can be destroyed
by interaction with the environment. To overcome this difficulty,
Bennett {\it etal.}, Deutsch {\it etal.} and Gisin {\it etal.}
presented several schemes to distill maximally entangled states of
two qubits out of a set of pairs in mixed entangled states[10-12].
These process are called entanglement distillation, which only
consists of local actions and classical communication.
Unfortunately, it is not known in general, which mixed state are
distillable at that time. Later on, the Horodecki family shown
that all entangled states of $2\times 2$ and $2\times 3$ systems
can be distilled into a singlet[13], and they proved that
nonpositivity of the partial transposition is a necessary
condition for the state of an arbitrary bipartite system to be
distillable[14]. But, it does not be the sufficient one. In fact,
there are states which have non-positive partial transpose[15] but
they are not distillable[16-18]. Recently, a useful sufficient
criterion, the so-called reduction criterion, has been
derived[18]. It shown that one can explicitly construct a protocol
to distill $\rho$ if there exists some vector $|\psi\rangle$
satisfying
\begin{equation}
\langle \psi|{\mbox Tr_B}\rho \otimes 1 -\rho|\psi\rangle <0,
\end{equation}
where ${\mbox Tr_B}$ stand for the trace over the second
subsystem. Moreover, some results of the distillability may be
generalized to the case of continuous variable systems[19].

Since the early study on the entanglement and distillability, most
works in this area are mainly concentrated on separability and
distillability for a concrete class of mixed states. In practice,
however, the destruction of a maximally entangled state is closely
related to a dynamical process. One of the examples is that the
destruction of a maximally entangled state due to the interaction
with the environment is a dynamical process. Some authors become
aware of the importance of the dynamical properties for quantum
entanglement[20,21].

In this paper we investigate the time evolution of the
distillability for a initially maximally entangled state. We
consider  a simple bipartite system which consists of two
particles with the same dimensions. Two cases are taken into
account in this paper. the first case is that only one of the two
particles is subjected to the environment, the other case is two
particles in the bipartite system are entirely under the effects
of the environment. From the viewpoint  of  dimensions, this paper
may be divided into two parts, i.e., $2\times 2$ systems and
$n\times n\ \ (n>2)$ systems. For the $2\times 2$ systems, there
is an alternative necessary and sufficient condition for
distillability.
 For high dimension, however, we only have
necessary or sufficient criterion for distillability, but not a
criterion for both. Our results show that the decoherence(caused
by the environment considered here)  do not change the
distillability of the $2\times 2$ systems, whereas the
distillability for a dissipation systems remains unchanged only
within a short time scales. For high dimension, however, the
distillability condition is more complicated. It does not only
depend on the coupling of the systems to the environment, but also
on the initial condition.

{\it Distillability.--}We consider two parties, Alice and Bob, who
share several pairs of particles. For simplicity, we assume here
that the particle has the same dimension $d$. Each pair is
initially in a maximally entangled state
\begin{equation}
|\psi^+\rangle=\frac{1}{\sqrt{d}}\sum_{i=1}^d|i,i\rangle,
\end{equation}
Interactions with the environment transform this pure state into
mixed state. This process may be described by a linear,
trace-preserving, completely positive map $L$ as
\begin{equation}
L(\rho_0)=\rho_f=\sum_{i=1}^k A_i \rho_0 A_i^{\dagger},
\end{equation}
where $\rho_0$ stands for the initial state, throughout this paper
we assume that $\rho_0=|\psi^+\rangle\langle \psi^+|$,i.e., the
system is initially in a maximally entangled state. The
trace-preserving property implies that the operator $A_i$ obey the
constraint
\begin{equation}
\sum_{i=1}^kA_i^{\dagger} A_i=1,
\end{equation}
with $1$ the identity matrix on the Hilbert space. One of the
physical implementations of the Kraus operators $A_i$ is as
follows. We consider a system interacting with the environment.
The evolution of the total system (system plus the environment) is
govern by a unitary operator $U(t)$. The reduced density matrix of
the system may be given by tracing the total density operator over
the environment,
\begin{equation}
\rho_f(t)=Tr_B[U(t)\rho_0(0)\otimes |0_B\rangle\langle 0_B|
U^{\dagger}(t)], \end{equation}
 where $|0_B\rangle$ stands for the
initial state of the environment. For a set of complete and
orthonormal bases $\{ \mu_B^i\}$ of the environment, one obtains
\begin{equation}
\rho_f(t)=\sum_i^k A_i\rho_0A_i^{\dagger},
\end{equation}
with $A_i=\langle \mu_B^i|U(t)|0_B\rangle$ is the so-called Kraus
operator. It is evident that $A_i$ satisfy $\sum_i^k A_i^{\dagger}
A_i=1$, for $U(t)$ is unitary.  So $L(\rho_0)\rightarrow
\rho_f(t)$ is a trace preserving completely positive map.

For $2\times 2$ systems, the sufficient and necessary condition
for distillability is
\begin{equation}
F_f\equiv {\mbox Tr}(\rho_0\rho_f)>\frac 1 2 ,
\end{equation}
In terms of the Kraus operators, this condition is
\begin{equation}
F_f=\sum_i^k|\langle\psi^+|A_i|\psi^+\rangle|^2>\frac 1 2 .
\end{equation}
In derivation of eq.(8), we used the initial condition
$\rho_0=|\psi^+\rangle\langle \psi^+|.$ For high dimension ($>2$),
there is a reduction criterion, it states that if there exists
some vector $|\psi\rangle$ such that
\begin{equation}
G_f\equiv \langle \psi|{\mbox Tr_B}\rho_f\otimes
1-\rho|\psi\rangle <0,
\end{equation}
then the final state $\rho_f$ is distillable. An important aspect
of this criterion is that if one finds a state $|\psi\rangle$
satisfying eq.(9), then one can explicitly construct a protocol to
distill $\rho _f$[18]. For a initially maximally entangled state
$|\psi^+\rangle$, we suppose that the most promising state, by
which the distillation proposal is constructed, is
$|\psi^+\rangle$ itself. In practice, we may always perform a
distillation before the maximally entangled state decoheres far
away from its initial state. In this sense the sufficient
condition for  distillability is
\begin{eqnarray}
G_f&=&\sum_{i=1}^k \langle \psi^+|{\mbox
Tr_B}(A_i|\psi^+\rangle\langle \psi^+|A_i^{\dagger})\otimes
1|\psi^+\rangle \nonumber\\ &<&\sum _{i=1}^k |\langle
\psi^+|A_i|\psi^+\rangle|^2.
\end{eqnarray}
Therefore if we get the Kraus operators, we may know exactly the
distillability of a state. Some words of caution are now in order.
The Kraus operators are not unique in general. For example, in
eq.(6) we may choose the other bases $\{\nu_B^i\}$ instead of
$\{\mu_B^i\}$ to compute the Kraus operators. However, different
sets of Kraus operators, which describe the same dynamical
process, may be transformed each to other by a unitary
transformation. In this sense, the distillability criterion do not
depend on the choice of Kraus operators.

{\it $2\times 2$ system with decoherence.--} Decoherence occurs
due to unwanted interactions between our quantum system and its
environment. These interactions cause only information  leak out
of the system. Typically, this process may be described by the
following two Kraus operators[22].
\begin{equation}
A_1=\left( \matrix{ 1 &0 \cr 0 & e^{-\gamma t}\cr } \right),\ \
A_2=\left(\matrix{ 0&0 \cr 0& \sqrt{1-e^{-2\gamma t}} \cr}\right
).
\end{equation}
We consider a pair of entangled particles a and b. If only one of
them (say a) are subjected to environment, the time evolution of a
maximally entangled state $$|\psi^+\rangle =\frac {1}{\sqrt 2}
(|0_a,1_b\rangle -|1_a,0_b\rangle )$$ is then given by
\begin{equation}
\rho_f(t)=\left(\matrix{0&0&0&0\cr 0&\frac 1 2 &\frac 1 2
e^{-\gamma_a t} &0\cr 0&\frac 1 2 e^{-\gamma_a t}& \frac 1 2 &
0\cr 0&0&0&0\cr} \right),
\end{equation}
where $\gamma_a$ is the decay rate for particle a. This process
may occur when a photon from a entangled pair is transmitted
through a fiber whose length is randomly modulated by acoustic
waves, or an atom from the pair is exposed to interactions with a
environment that consists of a set of harmonic oscillators[24].
Substituting eq.(11) into eq.(8), we obtain
\begin{equation}
F_f=\frac 1 2 +\frac 1 2 e^{-\gamma_a t}.
\end{equation}
It is evident that the final state is always distillable. If the
two particles are entirely under the effect of the environment, it
is easy to show that the new Kraus operator for the whole system
are
\begin{equation}
A_1= A_{1a}\otimes A_{1b},\ \ A_2=A_{2a}\otimes A_{2b},
\end{equation}
So, $F_f$ in this case is
\begin{equation}
F_f=\frac 1 2 +\frac 1 2 e^{-(\gamma_a+\gamma_b)t},
\end{equation}
the final state is always distillable, too.

{\it $2\times 2$ system with dissipation.--}
Different from the
case of decoherence, dissipation leads not only to the decay of
off-diagonal elements of the density matrix, but also the energy
loss (decay of the diagonal element of the density matrix). The
effect of energy loss to the environment is usually described by a
master equation[25]. which in the Born-Markov approximation may be
represented in terms of Kraus operators,
\begin{equation}
A_1=\left( \matrix{ 1 &0 \cr 0 & \sqrt{e^{-\lambda t}}\cr }
\right),\ \ A_2=\left(\matrix{ 0&-\sqrt{1-e^{-\gamma t}} \cr 0& 0
\cr}\right ).
\end{equation}
If we only transmit one of the entangled particles through a noisy
channel, the $F_f$ is then given by
\begin{equation}
F_f=\sum_{i=1}^2 |\langle \psi^+|A_i|\psi^+\rangle|^2=e^{-\frac
{\gamma_a t}{2}}.
\end{equation}
It is greater than $\frac 1 2 $ only for $t<2 ln2/\gamma _a$. This
indicates that we must perform distillation within $2ln2/\gamma_a$
in order to distill a maximally entangled particles. Similarly, we
obtain $F_f=e^{-\frac{(\gamma_a+\gamma_b)t}{2}}$ for the case that
the two entangled particles are both under the effects of the
environment. It is well known that the character time of
decoherence is much shorter than the dissipation one, so we can
ignore the effects of dissipation in general. From the viewpoint
of distillation, however, the dissipation is more destructive, for
the information loss due to decoherence may be reconstructed by
distillation, but the information loss caused by dissipation does
not. Figure 1 shows a results of $F_f(t)$ defined in eq.(7) by
solving master equations
$$\dot{\rho}=-i[H_0,\rho]+\frac{\gamma}{2}\sum_{i=a,b}(2\sigma_i^-\rho
\sigma_i^+-\sigma_i^+\sigma_i^-\rho-\rho \sigma_i^+\sigma^-_i),$$
and
$$\dot{\rho}=-i[H_0,\rho]+\frac{\gamma}{2}\sum_{i=a,b}(2\sigma_i^z\rho
\sigma_i^z-\sigma_i^z\sigma_i^z\rho-\rho \sigma_i^z\sigma^z_i)$$
numerically, where $H_0$ represents the free Hamiltonian and is
defined as $H_0=\omega(\sigma_a^z+\sigma_b^z)$. It is well known
that the first master equation describes the dynamics for a
dissipation system, while the second one governs a decoherence
process. Both master equations can be derived by using Markov
approximation and assuming system-bath interaction
$\sum_i(\sigma^+ a_i+h.c.)$ and $\sum_i(\sigma^z a_i+h.c.)$,
respectively. Here we use a notation $a_i$ to denote the bath mode
annihilation operator. It is clear that $F_f(t)$ corresponding a
dissipation process goes down below 0.5 earlier than a decoherence
process. In this sense, we say dissipation is more destructive.
\begin{figure}
\epsfxsize=7cm \centerline{\epsffile{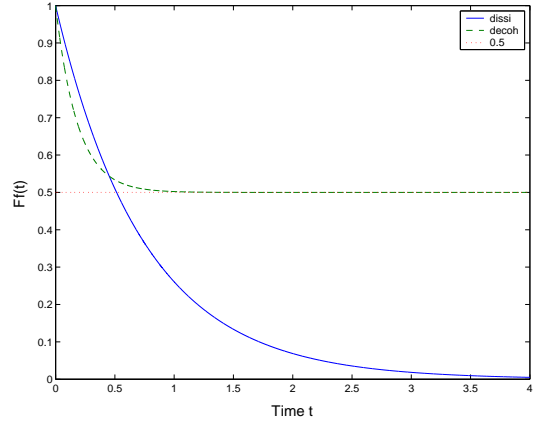}} \vskip 0.4cm
\caption[]{$F_f(t)$ versus t, the dashed line is drawn  for a
decoherence system, while the solid line is for a dissipation
system. Parameters chosen are $\gamma=0.6 \omega$, $t$ is chosen
in units of $1/\omega$. }
\end{figure}
 $n \times n$ {\it system with decoherence.--} Without  loss of
generality, we present the following model to describe the effect
of decoherence, the model Hamiltonian is
\begin{eqnarray}
H&=&H_s+H_B+H_i,\nonumber\\ H_i&=&\sum_jf_j(\{o_s\})\otimes
g_j(\{o_B\}),
\end{eqnarray}
where $f_j$ and $g_j$ are functions of system operators and bath
operators, respectively. In order to ensure that only the
decoherence occur in the system, we make a constraint
$[o_s,H_s]=0$ on the operator $o_s$. This condition implies that
we may factorize the time evolution operator of the composite
system in the following way
\begin{equation}
U(t)=e^{-i(H_s+H_B)t}\cdot\prod_{j=1}^N e^{F_j(t)\cdot
f_j(\{o_s\})\cdot G_j(\{o_B\})}.
\end{equation}
Here, $$ \{ H_s, H_B, f_j( \{o_s\})\cdot
G_j(\{o_B\})(j=1,2,...,N)\}$$ are elements of Lie algebra enlarged
by $H_s, H_B, H_i$ , while function $F_j(t)$ is determined by
\begin{eqnarray}
&\ &\frac{d}{dt}e^{F_j(t)\cdot f_j(\{o_s\})\cdot
G_j(\{o_B\})}=\nonumber\\ &=&\dot{F}_j(t) F_j(t)f_j(\{o_s\})
G(\{o_B\}) e^{F_j(t)\cdot f_j(\{o_s\})\cdot G_j(\{o_B\})}.
\end{eqnarray}
Eq.(20) indicates that in bases spanned by the eigenstates of
$H_s$, the diagonal elements of the system density matrix remain
unchanged in the evolution process, while the off-diagonal parts
gain a complex phase with the evolution. A simple calculation give
the element of reduced density matrix for the system
\begin{eqnarray} \langle m|\rho_s|n\rangle&=&{\mbox
Tr_B}\langle m|\rho|n\rangle \nonumber\\ &=&{\mbox Tr_B} \langle
m|U(t)\rho_s(0)\otimes\rho_B(0) U^{\dagger}(t)|n\rangle\nonumber\\
&=& {\mbox Tr_B} [
e^{-i(E_mt-E_nt)}\prod_{j=1}^Ne^{F_j(t)\bar{f}_{jm}G_j(\{o_B\})}\nonumber\\
&\cdot &\rho_{mn}(0)\otimes\rho_B(0)
e^{F^*_j(t)\bar{f}_{jn}G_j(\{o_B\})}]\nonumber\\ &\equiv&
\rho_{mn}(0)e^{-\gamma_{mn}(t)-i\Gamma_{mn}(t)},
\end{eqnarray}
where $\bar{f}_{jm}$ satisfies
$f_j(\{o_s\})|m\rangle=\bar{f}_{jm}|m\rangle$.
 The quantities $\gamma_{mn}(t)$ have
the following properties, $$\gamma_{mn}(t)=0,{\mbox for}\ \  m=n,
\ \ {\mbox and } \ \ \gamma_{mn}(t)>0$$ for otherwise. Physically,
no energy transfer between the system and the environment require
$\gamma_{mn}=0$ for $m=n$, while $\gamma_{mn}(t) >0$ for $m\neq
n$. For $\Gamma_{mn}(t)$, we have
$\Gamma_{mn}(t)=-\Gamma^*_{nm}(t)$, this property is directly from
the hermity of $\rho_f$. For a initially maximally entangled state
\begin{equation}
\rho_0=\frac 1 d\sum_{m,n}|m_a,m_b\rangle\langle n_a,n_b|,
\end{equation}
It is easy to check that
\begin{eqnarray}
G_f(t)&=&{\mbox Tr} ({\mbox Tr_b} \rho_f\otimes
1-\rho_f)=-\sum_{m,n}^d \frac 1 d
e^{-\gamma_{mn}^a(t)-\gamma_{mn}^b(t)}\nonumber\\ &\cdot &
\cos(\Gamma_{mn}^a(t)+\Gamma_{mn}^b(t)).
\end{eqnarray}
Here, we assume the two entangled particles are both under the
effect of the environment, $\gamma^a_{mn}(t)$ and
$\Gamma^a_{mn}(t)$ ($\gamma^b_{mn}(t)$ and $\Gamma^b_{mn}(t)$) are
defined by eq.(21) for a-particle(b-particle). Eq.(23) is a
damping-oscillation function of time. If
$\cos(\Gamma_{mn}^a(t)+\Gamma^b_{mn}(t))>0$, the final state of
$\rho_f$ is distillable. If $\rho_f$ does not violate the Peres
separability  criterion, the final state $\rho_f$ can not be
distilled, this can be done for a given state $\rho_f$. For a
model presented in [23], $\Gamma_{mn}^x(t), (x=a,b)$ may be
rewritten explicitly as
\begin{equation}
\Gamma_{mn}^x(t)=\int
\frac{g_{\omega,x}^2}{\omega^2}(m^2-n^2)(\omega t-\sin\omega
t)\rho(\omega) d\omega,
\end{equation}
where $g_{\omega,x}$ is the coupling constant of particle $x$ to
the environment, $\rho(\omega)$ stands for the spectrum
distribution of the environment. Eq.(23) shows  that the final
state $\rho_f$ is not always distillable, its distillability would
depend on the detailed information of the system even in the case
of decoherence.

 $n \times n$ {\it system with dissipation.--} We consider a spin-n
($d=2n+1)$ particle interacting with its environment. Under the
Born-Markov approximation, the dissipation process may be
described by a master equation in the Lindblad form[25]
\begin{equation}
\dot{\rho}=-i[H_0,\rho]+\frac{\gamma}{2}(2S_-\rho
S_+-S_+S_-\rho-\rho S_+S_-),
\end{equation}
where $S_+(S_-)$ is spin operator defined by $S_+=(S_x+iS_y)$
($S_-=(S_x-iS_y)$). $H_0=\Omega S_z$ is the free Hamiltonian of
the spin-n particle. $\gamma$ is the decay rate. We would like to
note that eq.(25) only described the time evolution of one
particle in the entangled pair. In general, the particles in the
pair experience different environment, for they are transmitted
trough different noisy channels. By using the method presented
in[26], we obtain the element of the density matrix up to the
first order of $\gamma$
\begin{eqnarray}
\rho_{mn}(t)&=&\frac 1 d +\frac{\gamma t
}{d}\sqrt{(f+m+1)(f-m)(f+n+1)(f-n)}\nonumber\\ &-& it\frac 1 d
(\Omega m+\Omega n-i\gamma (f+m)(f-m+1)\nonumber\\ &-&i\gamma
(f+n)(f-n+1)).
\end{eqnarray}
In derivation of eq.(26), the initial condition $\rho(0)=\frac 1 d
\sum_{m,n}|m\rangle\langle n|$ was used. If the two particles are
both exposed to the interactions of the environment, following the
procedure presented in case A of this section, we arrive at
\begin{eqnarray}
G_f(t)&=&-\frac{1}{d^2}\sum_{m>n}^d 2{\mbox Re} [1+\gamma
t\sqrt{(f+m+1)(f-m)}\nonumber\\ &\cdot&
\sqrt{(f+n+1)(f-n)}\nonumber\\ &-& it (\Omega m+\Omega n-i\gamma
(f+m)(f-m+1)\nonumber\\ &-&i\gamma (f+n)(f-n+1))]^2.
\end{eqnarray}
According to the sufficient condition(8), the final state is
distillable if $G_f(t)<0$. For  $\gamma=0.2\Omega$, the dependence
of the critical time $t_c$ defined by $G_f(t_c)=0$  on dimension
of the systme is shown in Fig.2.  As Figure 2 shows,  the larger
the dimension of the system, the shorter the character time of
distillability destruction.

\begin{figure}\label{fig:potential}
\epsfxsize=7cm \centerline{\epsffile{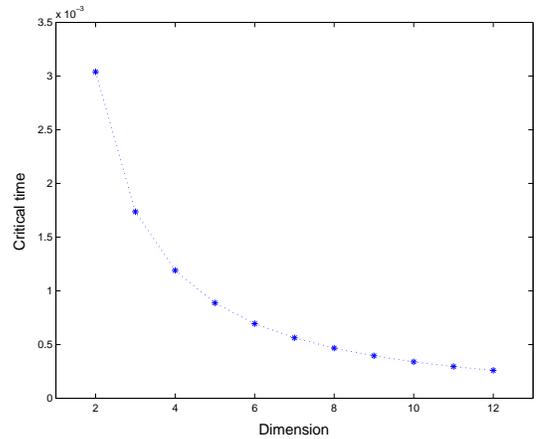}} \vskip 0.3cm
\caption[]{The dependance of the critical time $t_c$ on system
dimension}
\end{figure}
Figure 3 is plotted for a numerical simulation of $G_f(t)$ for a
decoherence system and a dissipation system, the dynamics of the
dissipation system is govern by master equation (25), while the
time evolution of the decoherence system are described by
\begin{equation}
\dot{\rho}=-i[H_0,\rho]+\frac{\gamma}{2}(2S_z\rho
S_z-S_zS_z\rho-\rho S_zS_z),
\end{equation}
\begin{figure}\label{fig:potential}
\epsfxsize=7.1cm \centerline{\epsffile{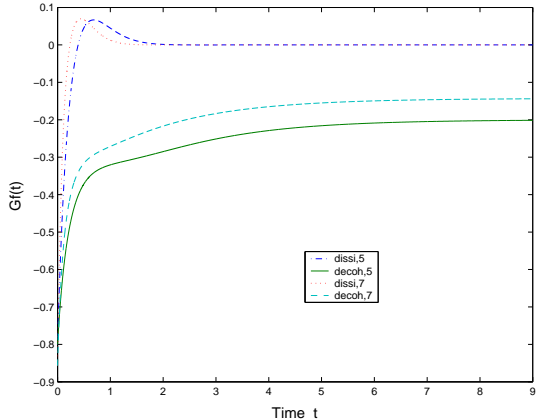}} \vskip 0.5cm
\caption[]{ $G_f(t)$ as a function of time.Solid line is for a
decohence system with dimension 5, dotted line for a dissipation
one with dimension 7, dashed line for decoherence with dimension 7
and dotted and dashed line for dissipation with dimension 5. In
this plot, we choose $\gamma=0.6\Omega$ and time $t$ is plotted in
units of $1/\Omega$ }
\end{figure}

From figure 3 we can see that in the case of dissipation $G_f(t)$
go up to zero after a short time evolution from our initial state
$|\psi(0)\rangle=\frac{1}{d}\sum_m|m,-m\rangle$, whereas $G_f(t)$
for  a decoherence process is always below zero. We would like to
address that this conclusion is not a general one, which would
depend on the initial state of the system and the bath
information, for example, if we choose $m=0$, and $\rho(\omega)=
constant$ in eq.(24), we may find a time when
$\cos(\Gamma_{mn}^a(t)+\Gamma_{mn}^b(t))<0.$

 In summary, the
dynamics of distillability for a bipartite system are investigated
in this paper. The destruction of the maximal entanglement for the
system is closely related to the interactions with the
environment. There are two kinds of interactions which lead to the
destruction of maximal entanglement. One is the quantum
decoherence, the another is the dissipation. From the viewpoint of
distillability, the dissipation is more harmful in quantum
communication. For example, the decoherence do not change the
distillability of a $2\times 2$ systems, whereas the dissipation
does. For high dimension, both dissipation and decoherence do
destroy the quantum entanglement, but in different ways.

We would like to note that the initial state considered here is a
maximally entangled state. If the initial state belongs to a
special class of entangled states, the local environment can
enhance the quantum entanglement[27] from the view point of
quantum teleportation. This increases the efficiency of
distillation and makes the undistillable state to be distillable .
In the framework of quantum information theory, the state change
allowed by quantum mechanics may be classified into three types.
The first one is the unitary evolution, the second is the
interaction with a environment, and the last one is a measurement
performed on the quantum system. The unitary evolution
 is of
cause change the state of the bipartite system, for the two
particles in the bipartite systems may interact each other.
Although we do not discuss the dynamic of distillability of such a
system in this paper, the method of this paper can easily
generalized to this case. As to the measurement, the method
presented here is also available, because the most general type of
measurement can be understood within the framework of unitary
evolution[28]. In fact most generalized measurement can be
realized through many dynamical processes[29].

{\bf \large ACKNOWLEDGEMENT:}\\ This work is supported by the NSF
of China.\\

\end{multicols}
\end{document}